\documentclass[reprint,aps,pra,twocolumn,groupedaddress]{revtex4-2}

\usepackage{amsmath,amssymb,amsfonts}
\usepackage{graphicx}
\usepackage{siunitx}
\usepackage{natbib}
\usepackage{braket}
\usepackage{multirow}
\usepackage[normalem]{ulem}
\usepackage{verbatim}

\usepackage{url}
\usepackage[breaklinks]{hyperref}
\hypersetup{
    unicode=true,
    colorlinks=true,
    linkcolor=blue,
    citecolor=blue,
    urlcolor=blue
  }
\usepackage{breakurl}
\usepackage{xcolor}

\begin{document}

\title{Generation of high winding-number superfluid circulation in Bose-Einstein condensates}

\author{Kali E. Wilson$^1$}
\thanks{Present address: Department of Physics, SUPA, University of
Strathclyde, Glasgow G4 0NG, UK}
\author{E. Carlo Samson$^1$}
\thanks{Present address: Department of Physics, Miami University, Oxford, OH, USA}
\author{Zachary L. Newman$^1$}
\thanks{Present address: Octave Photonics, Louisville, CO 80027, USA}
\author{Brian P. Anderson$^1$}
\email{bpa@optics.arizona.edu}
\affiliation{$^1$Wyant College of Optical Sciences, University of Arizona, Tucson, AZ 85721, USA}

\date{\today}

\begin{abstract}
We experimentally and numerically demonstrate a method to deterministically generate multiply-quantized superfluid circulation about an obstacle in highly oblate Bose-Einstein condensates (BECs).  Our method involves spiraling a blue-detuned  laser beam, which acts as a repulsive optical potential, around and towards the center of the BEC.  This optical potential serves first as a repulsive stirrer to initiate superflow within the BEC, and then as a pinning potential that transports the center of the superfluid circulation to the center of the condensate.  By  changing  the  rate  at  which  the  beam  moves  along  the spiral  trajectory,  we  selectively  control  the  net circulation  introduced  into  the  BEC.  We experimentally achieve pinned superflow with winding numbers as high as 11, which persists for at least 4\,s. 
At the end of the spiral trajectory, with the pinning beam on at full power, the BEC has a toroidal geometry with a high winding-number persistent current.  Alternatively, the beam power can be ramped off, allowing controlled placement of a cluster of singly-quantized  vortices of the same circulation. This technique can serve as a building block in future experimental architectures to create on-demand vortex distributions and superfluid circulation in BECs.  
\end{abstract}

\maketitle
\section{Introduction}
Highly-oblate dilute-gas Bose-Einstein condensates (BECs) \cite{Pethick2008}, in which fluid dynamics occur primarily in two-dimensions, have opened up experimental studies of two-dimensional (2D) quantum turbulence \cite{Neely2013, Wilson2013, Gauthier2019, Johnstone2019}, as well as theoretical and numerical studies of point vortex models and the complex collective behaviour of a distribution of many vortices \cite{Bradley2012, Reeves2020, Stockdale2020}.  In these scenarios, the initial placement of the vortex cores determines the system's quantum phase profile and hence the subsequent fluid flow and vortex dynamics, and vortex behaviour can highlight fundamental differences between superfluids and classical viscous flows \cite{Musser2019, Stockdale2020}.  Yet interest in vortex dynamics extends well beyond basic aspects of superfluidity.  Vortex dynamics also play a role in analog cosmology \cite{Eckel2018, Banik2021}, where large-quanta vortices are used to mimic rotating black holes \cite{Geelmuyden2021} and to study ergoregion instabilities \cite{Giacomelli2020}. As experiments progress to include quantum mixtures and binary superfluid dynamics \cite{Ferrier2014, Yao2016, Lee2018, Wilson2021}, vortices once again become highly relevant as probes of the macroscopic quantum state \cite{Kuopanportti2019}.  Of particular recent interest is whether vortices survive in quantum-fluctuation enhanced regimes such as the Lee-Huang-Yang gas \cite{Lee1957, Skov2021}, or quantum droplets \cite{Cabrera2018, Semeghini2018, Kartashov2018, Tengstrand2019}.  These states are formed in quantum mixtures where the net mean-field interaction is tuned closed to zero such that beyond-mean-field effects like quantum fluctuations play an enhanced role in governing the system's behavior. 

Exploring such superfluid physics with BECs requires developing experimental techniques for deterministic vortex generation with control over placement, vorticity, and direction of circulation. This is particularly relevant for studies of quantum turbulence where one might want to reproducibly generate a many-vortex state to look for signatures of quantum turbulence in subsequent vortex dynamics. Progress towards a flexible experimental vortex architecture has been made in highly-oblate single-component BECs regarding deterministic placement of individual vortex cores \cite{Samson2016}, with subsequent refinement made possible by the development of arbitrary configurable optical potentials \cite{Gauthier2016, Kwon2021}. However, a fully flexible architecture would benefit from the further development of additional experimental techniques that allow for controlled placement of a cluster of a fixed number of vortices, all with the same sign of circulation.  
In this paper, we present a controlled vortex generation method that can generate large net superfluid circulation and multiply-charged vortex states with observed winding numbers up to 11. The vortices are pinned to the beam and can be moved to a desired location within the BEC, or can be released from the beam for studies of vortex dynamics. 

Various methods have been proposed and used to create quantized vortices in BECs. Early techniques included density and phase-engineering in a two-component condensate \cite{Matthews1999}, and rotating the confining potential \cite{Madison2000a, Abo2001, Hod2001.PRL88.010405, Hal2001.PRL87.210403}. 
In the absence of pinning potentials, such rotation leads to the formation of a vortex lattice of singly-quantized vortices all with the same sign of circulation, however, aggregation of vortices into one giant circulation has been achieved by applying a focused laser beam at the center of the rotating BEC \cite{Eng2003.PRL90.170405}.   Multiply-charged vortex states can also be created through topological phase imprinting methods \cite{Lea2002.PRL89.190403, Shi2004.PRL93.160406, Kum2006.LP16.371, Okano2007}, in which winding numbers of two and four were obtained. 
In general, unpinned multiply-quantized vortices are unstable in single-component BECs, and tend to break apart into singly-quantized vortices.  However, persistent currents have been demonstrated in toroidal geometries where the presence of a central pinning potential keeps the current from dissociating into individual vortex cores \cite{Ryu2007, Wright2013, Neely2013}.  Much of this early work focused on introducing vorticity into the BEC, but did not focus on controlled placement of individual vortices or large-net-vorticity clusters. More recently, digital micromirror devices combined with high numerical aperture objectives have enabled optical traps and stirring potentials with greater resolution and enhanced dynamic control over vortex generation and placement \cite{Gauthier2016, Gauthier2019,Johnstone2019}. Thus the development of a wide variety of techniques for controlling winding number and vortex cluster placement continues to be highly relevant and desirable.

The paper is organized as follows.  In Sec.~\ref{sec:concept}, we introduce the conceptual foundation for our technique.  In Sec.~\ref{sec:exp} we describe the details of our experimental studies and observations.  In Sec.~\ref{sec:sims} we present results from corresponding simulations of the 2D Gross-Pitaevskii equation that illuminate the process for creating pinned superfluid circulation, or multiply-quantized vortices. In Sec.~\ref{sec:winding} we explore the relationship between stirring speed and the number of pinned vortices. In Sec.~\ref{sec:twobeams} we discuss prospects for extending the technique to multiple stirring beams and hard-wall trapping potentials. Section~\ref{sec:conclusion} concludes the article.

\section{Concept}\label{sec:concept}
%
\begin{figure}[t!]
\includegraphics[width = \columnwidth]{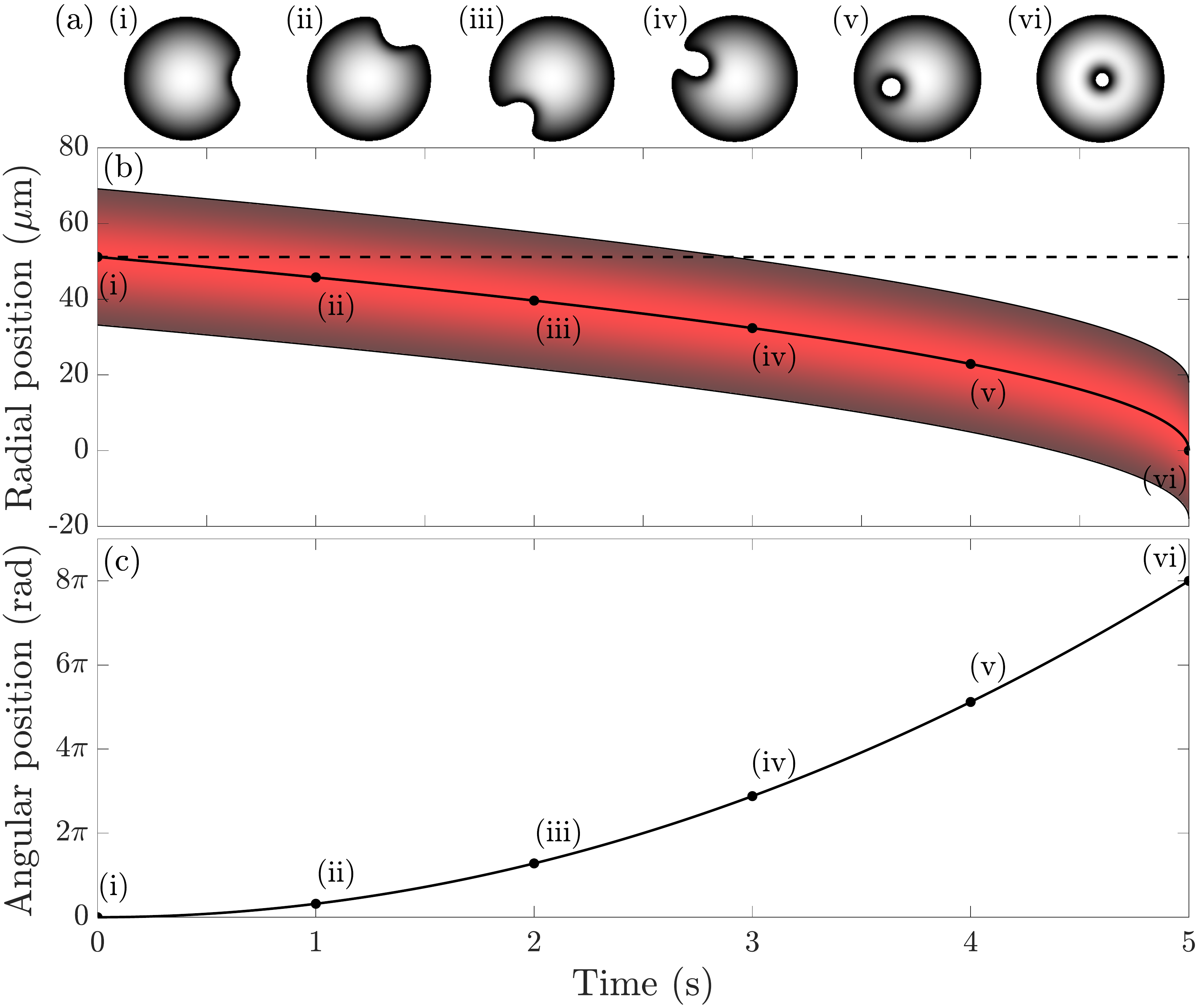}
\caption{Spiral beam trajectory $\vec{s}(t) = \{r(t), \theta(t)\}$, with radial position $r(t) = R_0\sqrt{1-t/\tau_\mathrm{s}}$, angular position $\theta(t) = 2\pi N_\mathrm{s} \left(t/\tau_\mathrm{s} \right)^2$ and spiral parameters $N_\mathrm{s} = 4$, $\tau_\mathrm{s} = 5~\mathrm{s}$, $R_0 = R_r$. See text for definitions of parameters. (a) Simulated 2D density profiles of the BEC in the presence of the beam are shown at 1-s intervals [density profiles (i)-(vi) correspond to the labeled time points in plots (b) and (c)]. (b) Radial beam position $r(t)$ as a function of time during the spiral trajectory. Red shading indicates the radial extent of the Gaussian beam. The horizontal dashed line indicates the radial Thomas-Fermi radius of the circular condensate, $R_r$. (b) Angular beam position $\theta(t)$ as a function of time during the spiral trajectory.}
\label{fig:concept}
\end{figure}

Our method involves spiraling a repulsive optical potential formed by a blue-detuned laser beam around and inwards towards the center of the BEC as depicted in Fig~\ref{fig:concept}.
This optical potential serves first as a repulsive stirrer to initiate circular superflow within the BEC, and then as a pinning potential that transports the center of the superfluid circulation to the center of the condensate. 
 
The spiral sequence begins with the blue-detuned laser beam placed just at the edge of our highly-oblate axially symmetric (about the $\hat{z}$ axis) BEC [see density profile (i) in Fig.~\ref{fig:concept}(a)]. This stirring beam acts as a repulsive obstacle that is spiraled inwards in the $\hat{x}$-$\hat{y}$ plane towards center of the BEC  at $x = y = 0$. Figure~\ref{fig:concept}(a) shows the simulated 2D density profile of the BEC at relevant time points within the spiral sequence. In this example the beam travels counterclockwise.  The radial and angular trajectories of the beam are shown in Figs.~\ref{fig:concept}(b) and \ref{fig:concept}(c), respectively. 
As the stirring beam initially spirals inwards, it pushes fluid out of the way and initiates local flow around the inner edge of the beam.  Given that the beam initially moves along the edge of the BEC, the induced superflow is uni-directional and the direction, either clockwise (CW) or counter-clockwise (CCW), is fixed by the direction of motion of the stirring beam.  Thus, a beam spiraling CCW with respect to the center of the BEC [see insets in Fig.~\ref{fig:concept}(a)], also initiates CCW local superflow around the beam. This is in contrast to the methods in Refs.~\cite{Neely2010, Samson2016, Staliunas2000} that nucleate vortex dipoles, pairs of vortices with equal but opposite-signed circulation. We note that the method proposed in Ref.~\cite{Staliunas2000} bears some similarity to our technique in that (after vortex dipole nucleation) a laser beam is used to pin one of the vortices and then guide it to the center of the BEC. However, due to the fundamental difference in the vortex creation process the method of Ref.~\cite{Staliunas2000} does not extend to multiply-quantized vortices.  

Eventually the beam moves sufficiently inwards that on the outside edge of the beam, the fluid behind the beam merges with the fluid in front of the beam, yielding connected circular flow around the beam [see density profile (v) in Fig.~\ref{fig:concept}(a)].  At this point circulation has been brought inside the BEC and pinned to the beam. The winding number associated with the pinned circulation is fixed by the velocity of the now-continuous superfluid flow.  Thus the winding number of the multiply-quantized circulation that is to be pinned to the beam can be controlled by varying the speed of the stirring beam.  For sufficiently high beam powers and low spiraling speeds the circulation remains pinned to the beam as the beam continues its trajectory to the center of the condensate. 
Once the pinning beam has reached the desired final position, it can be left in place at full power such that the condensate is in a toroidal geometry with a high-winding number persistent current \cite{Ryu2007}.  Alternatively, the beam's power can be ramped off, allowing placement of a cluster of singly-quantized vortices of the same sign of circulation, with control over the placement of the cluster's centroid within the BEC.  
 
The particular trajectory $\vec{s}(t) = \{r(t),\theta(t)\}$ used in our experiment was chosen to mimic the variation of the speed of sound in the condensate, i.e., low at the edges and highest at the center. Here $r(t) = R_0\sqrt{1-t/\tau_\mathrm{s}}$ plotted in Fig.~\ref{fig:concept}(b) is the radial distance of the beam from the center of the BEC,  and $\theta(t) = 2\pi N_\mathrm{s} \left(t/\tau_\mathrm{s} \right)^2$ plotted in Fig.~\ref{fig:concept}(c) is its angular displacement with respect to the axis defined by the center of the BEC and the initial beam position. Here $R_0$ is the initial radial position of the beam, $N_\mathrm{s}$ is the number of $360^{\circ}$ rotations within the spiral, and $\tau_\mathrm{s}$ is the total time duration of the spiral trajectory. The speed of the circular superflow initiated by the beam, and the associated winding number are fixed by the spiral parameters $N_\mathrm{s}$ and $\tau_\mathrm{s}$.  We emphasize that the exact nature of the trajectory and the diameter of the stirring beam are not critical for the success of the method so long as the initial motion of the beam is approximately tangent to the outer edge of the condensate. However, it is important to manage the speed of the beam to (1) allow the fluid ahead and behind the beam to merge in a controlled fashion to avoid generating excitations such as dark solitons \cite{Burger1999, Denschlag2000, Anderson2001}, (2) stay below the critical speed for dipole nucleation \cite{Neely2010} once the beam has fully entered the condensate, and (3) allow the multiply-quantized circulation to remain pinned to the beam as it moves within the condensate.  

\section{Experiment}\label{sec:exp}
%
\begin{figure}[t!]
\includegraphics[width = \columnwidth]{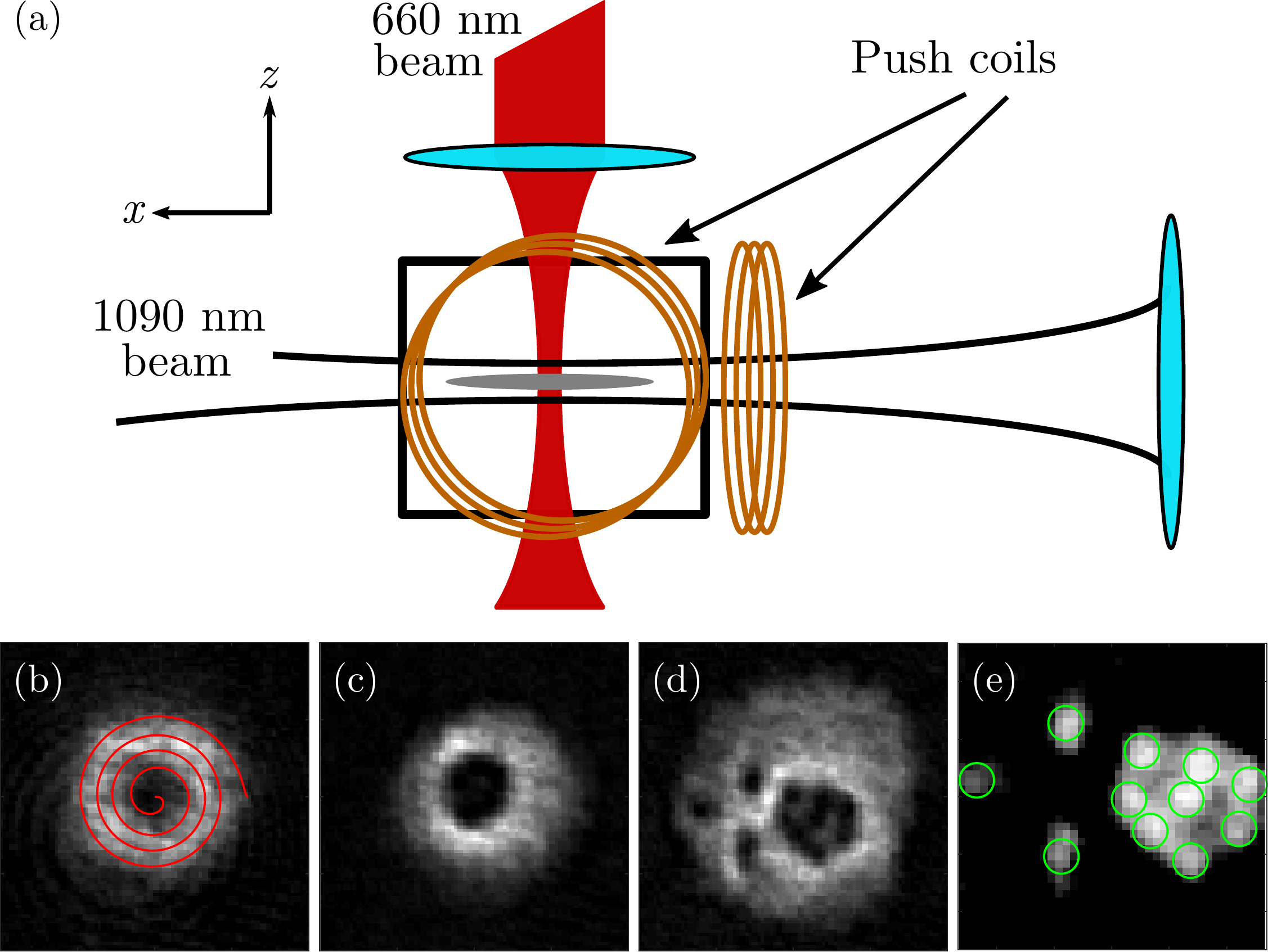}
\caption{Experimental setup. (a) Schematic of the experimental setup (not to scale). (b)-(d) Axial absorption images showing a $200$-$\mu\mathrm{m}$ square field of view in the horizontal $x$-$y$ plane. (b) In-trap image of the BEC with the 660-nm blue-detuned beam at its center.  Also shown is the effective spiral trajectory [red line] of the 660-nm beam as the condensate is translated by time-varying magnetic fields. (c) BEC after ballistic expansion.  The BEC is allowed to expand right after the pinning optical potential is ramped off ($t_d = 0~\mathrm{ms}$, see text).  The large hole in the center of the condensate is taken as evidence of multiply-charged superflow since its size is much larger than a single vortex core. (d) BEC with 11 singly-quantized vortices, imaged after ballistic expansion. Here we wait for $t_d = 160~\mathrm{ms}$ between the 660-nm beam ramp off and ballistic expansion to allow the superflow to disperse into singly-quantized vortices. (e) Residuals from Thomas-Fermi fit to the image shown in (d), zoomed in to the central vortex region. Green circles indicate individual vortex cores.}
\label{fig:exp}
\end{figure}

We create highly-oblate BECs of $^{87}$Rb in the $5^2S_{1/2} \vert F = 1, m_F = -1 \rangle$ hyperfine state, confined in a  hybrid magnetic-optical harmonic trap.  Radial confinement in the $\hat{x}$-$\hat{y}$ plane is provided by a time-averaged orbiting potential (TOP) magnetic trap \cite{Petrich1995}, with the axis of symmetry along the vertical ($z$) direction.  The vertical confinement from the TOP trap is enhanced by a red-detuned laser sheet at 1090 nm, which propagates along the $\hat{x}$ direction and is tightly focused in the $\hat{z}$ direction \cite{Neely2010, Neely2013, Samson2016} as shown in Fig.~\ref{fig:exp}(a).  In the absence of any additional optical potentials, the hybrid trap has radial and vertical trap frequencies of $\left(\omega_r,\omega_z \right) = 2\pi \times (8,90)$ Hz, respectively.  The trapped BECs have typical atom numbers of $N_\mathrm{c} \sim 2 \times 10^6$, a chemical potential of $\mu_0 \sim 8 \hbar \omega_z$, and radial Thomas-Fermi radii of $R_r \sim 50 \, \mu\mathrm{m}$.

For the spiral technique presented here, we employ an additional focused blue-detuned laser beam at 660 nm, which penetrates the condensate as shown in Fig.~\ref{fig:exp}(a). The 660-nm beam propagates along $\hat{z}$ with a focused Gaussian $1/e^2$ radius of $w_0 \sim18 \, \mu\mathrm{m}$ at the location of the condensate. Figure~\ref{fig:exp}(b) shows a representative \emph{in-situ} absorption image of a BEC with the blue-detuned beam centered on the BEC (the final beam position). 
In our experimental configuration the blue-detuned beam is stationary with respect to the laboratory rest frame, and time-varying magnetic fields [`push coils' in Fig.~\ref{fig:exp}(a)] are used to move the BEC with respect to the beam.
In a typical experimental sequence, we form a BEC in the hybrid 1090-nm + magnetic TOP trap.  We use the push coils to translate the BEC to the initial spiral position over $\sim 1\,\mathrm{s}$.  We ramp on the 660-nm beam over $\sim 500\,\mathrm{ms}$ resulting in our initial spiral configuration with the spiral beam located at $r(t = 0) \sim R_r$ [see density profile (i) in Fig.~\ref{fig:concept}(a)]. We then execute the spiral trajectory indicated by the red line in Fig.~\ref{fig:exp}(b).  As discussed in Sec~\ref{sec:concept}, the position of the focused beam in the rest frame of the BEC is described by $\vec{s}(t) = \{r(t),\theta(t)\}$, with $r(t) = R_0\sqrt{1-t/\tau_\mathrm{s}}$,  and $\theta(t) = 2\pi N_\mathrm{s} \left(t/\tau_\mathrm{s} \right)^2$.

After a subsequent hold time $t_h \sim 4 \, \mathrm{s}$ (up to 7 s), we slowly ramp off the power of the blue-detuned beam ($t_\mathrm{ramp} \sim 0.5 - 1 \, \mathrm{s}$), and then remove the other trapping potentials, letting the BEC undergo a period of ballistic expansion prior to imaging. $t_\mathrm{ramp}$ is determined experimentally with the criteria of being long enough to avoid exciting the BEC, but short enough so that the multiply-quantized vortex pinned to the beam does not have time to disperse appreciably.  Figure~\ref{fig:exp}(c) shows a representative absorption image of the condensate immediately following the blue-detuned beam ramp off and subsequent 56-ms of ballistic expansion.  The large central region devoid of atoms in the center of the expanded BEC is much larger than what we expect for a singly-quantized vortex core.  Instead, the giant hole is indicative of a multiply-quantized vortex in the expanding BEC that was created by the spiraling process and then pinned to the blue-detuned beam \cite{Wright2013}.

%
\begin{figure}[t!]
\includegraphics[width=\columnwidth]{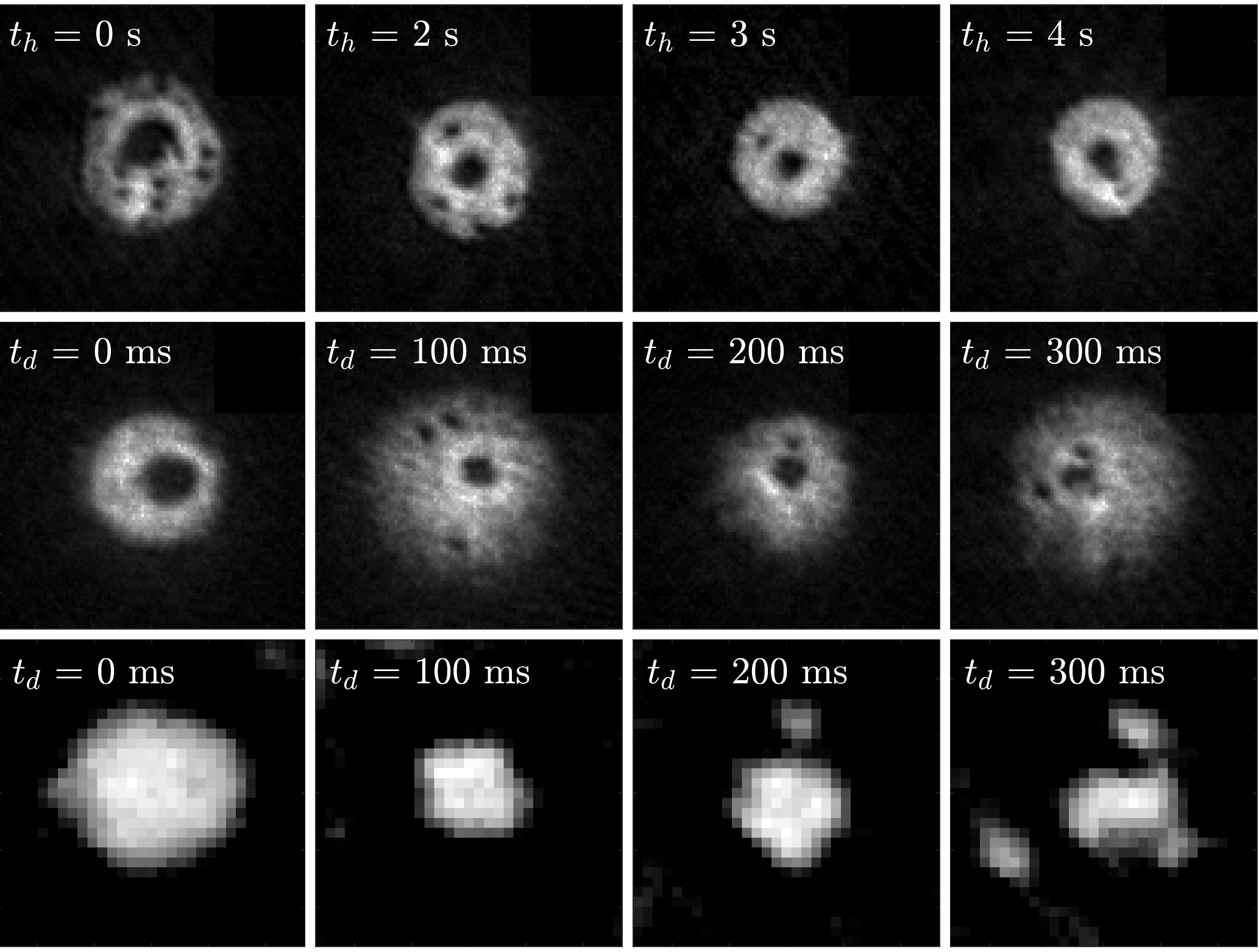}
\caption{BEC expansion images for variable hold time $t_h$ and superflow dispersal time $t_d$. The 200-$\mu$m square axial absorption images (upper and middle row) are taken after a period of ballistic expansion.  Top row: variable hold time, prior to 600-ms beam ramp off. Spiral parameters are $N_\mathrm{s} = 4$, $\tau_\mathrm{s} = 4.8~\mathrm{s}$. Here the BEC is allowed to expand directly after ramping off the beam ($t_d = 0$ ms).  Middle row: variable dispersal time $t_d$, after the 600-ms beam ramp off. Spiral parameters are $N_\mathrm{s} = 5$, $\tau_\mathrm{s} = 3.5~\mathrm{s}$. Here we hold the BEC for $t_h = 4~\mathrm{s}$, ramp the beam off, and then hold for a variable dispersal time $t_d$ to allow the superflow to dissociate into singly-quantized vortices.  Bottom row: residuals after fitting the images in the middle row to a Thomas-Fermi density profile, zoomed in to the central vortex region.} 
\label{fig:time}
\end{figure}  

To determine the winding number of the pinned superflow, we add an additional short hold time of $t_\mathrm{d} \sim 150 - 250 \, \mathrm{ms}$ after ramping off the blue-detuned beam, and prior to ballistic expansion.  As shown in Fig.~\ref{fig:exp}(d), adding the additional dispersal time enables us to resolve the individual vortex cores as the large multiply-quantized vortex core dissociates into singly-quantized vortices.  Figure~\ref{fig:exp}(d) shows a representative high-winding number superflow after it has dissociated into 11 individual vortex cores.  In Fig.~\ref{fig:exp}(e) we show residuals from a Thomas-Fermi fit to the 11-vortex image shown in Fig.~\ref{fig:exp}(d), zoomed-in to the central vortex region. Here green circles indicate individual vortex cores.

Figure~\ref{fig:time} shows representative images at the end of the spiral trajectory for varying $t_\mathrm{h}$ (top row, $t_\mathrm{d} = 0\,\mathrm{ms}$), and varying $t_\mathrm{d}$ (middle row, $t_\mathrm{h} = 4\,\mathrm{s}$).  The bottom row shows residuals  from a Thomas-Fermi fit to the corresponding images shown in the middle row, zoomed in to the central vortex region.  We first discuss the role of the hold time prior to the beam ramp off. For $t_\mathrm{h} = 0\,\mathrm{s}$, we observe both a central current and a number of unpinned vortices. Unpinned cores may be the result of too high of a beam velocity during the spiral, resulting in vortices depinning from the beam.  Or they may simply be a result of introducing more vorticity than can be stably pinned to the beam \cite{Law2014}. As we increase $t_\mathrm{h}$ we find fewer unpinned vortices until finally for hold times $t_\mathrm{h} = 3.5 - 4.0 \,\mathrm{s}$ we observe just the central pinned multiply-quantized vortex and no unpinned cores.
The latter likely leave the condensate due to their interaction with the thermal background \cite{Staliunas2000, Rooney2010}. Images of the expanded BEC for short hold times $t_h = 500 \, \mathrm{ms}$ show that there are on average 6 unpinned vortices for $\tau_\mathrm{s}=5$ s ($N_\mathrm{s} = 4$) and 4 unpinned vortices for $\tau_\mathrm{s} = 7$ s ($N_\mathrm{s} = 4$).

In the absence of a pinning potential (such as our blue-detuned beam), a multiply-charged vortex state is unstable \cite{Castin1999, Garcia1999, Butts1999, Simula2002}, and will tend to dissociate into individual vortices \cite{Simula2002}, which then disperse outwards from the center of the condensate \cite{Stockdale2020}. As shown in the middle row of Fig.~\ref{fig:time}, the vortex cluster begins to dissociate \emph{after} we ramp off the pinning beam.  Here we hold the BEC for $t_h = 4\,\mathrm{s}$, ramp the beam off over 600 ms, and then hold for a variable dispersal time $t_d$ to allow the multiply quantized superflow to dissociate into singly-quantized vortices.  The leftmost image $t_\mathrm{d} = 0 \, \mathrm{ms}$ shows the multiply-quantized vortex prior to dissociation.  As we increase $t_\mathrm{d}$, we observe the central superflow break apart into 4-5 vortices, with individual singly-quantized vortices resolvable around $t_\mathrm{d} = 200 \, \mathrm{ms}$.  To aid in resolving the individual vortex cores at short dispersal times, the bottom row of Fig.~\ref{fig:time} shows the corresponding residuals after fitting the images in the middle row to a Thomas-Fermi density profile.  The high variance in the experimental data (see below) makes it difficult to take a consistent time series.  However, we do observe that the disassociation of the cluster is not instantaneous, consistent with Ref.~\cite{Stockdale2020}.

\section{Numerical simulations}\label{sec:sims}
\begin{figure}[t!]
\includegraphics[width=\columnwidth]{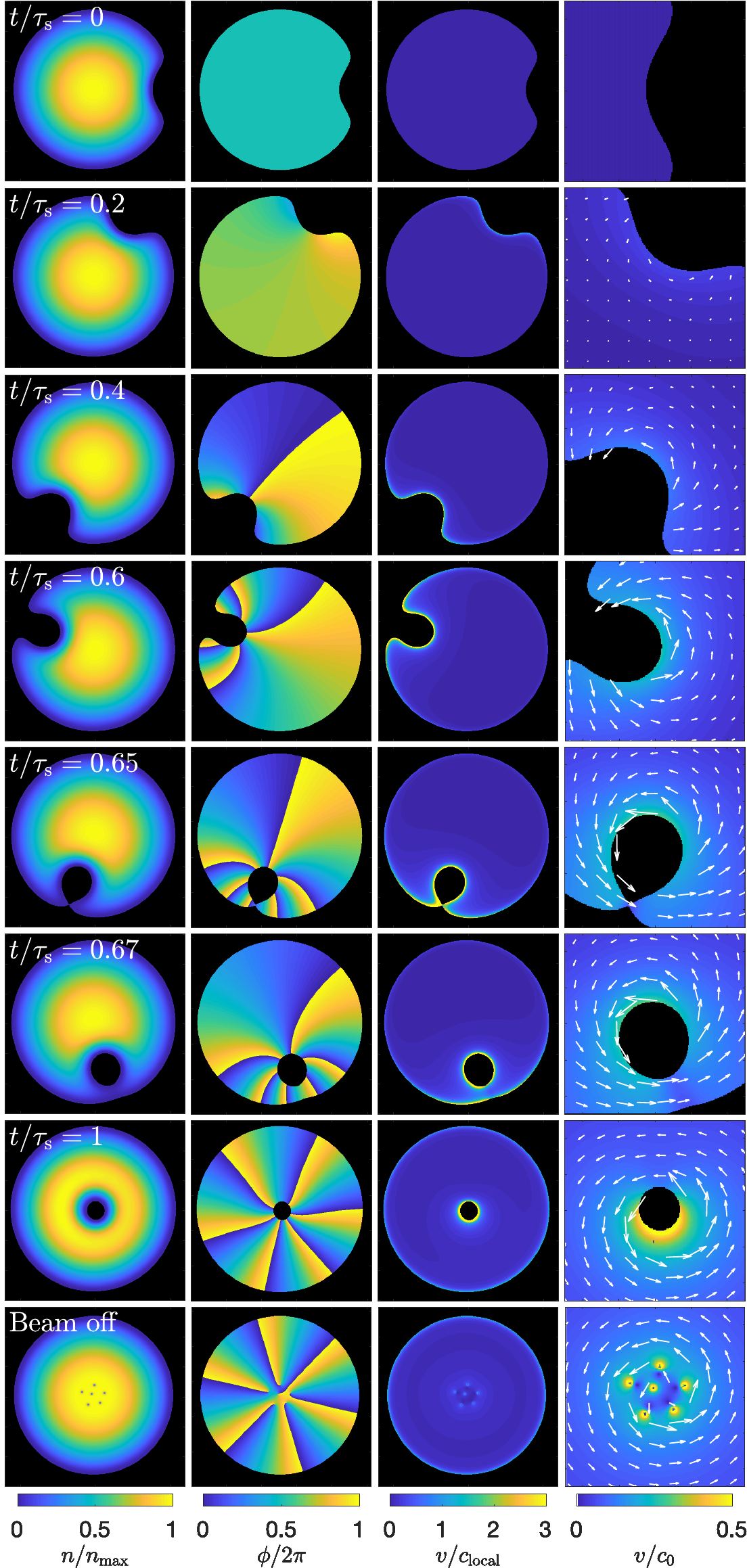}
\caption{Snapshots from 2D GPE simulation of the spiral technique at representative times $t/\tau_\mathrm{s}$. Columns from left to right show snapshots of (i) density, (ii) phase, (iii) the velocity profile scaled to the local speed of sound, and (iv) a zoomed in region of the velocity profile at the location of the stirring beam. The velocity profile in column (iv) is scaled to the bulk speed of sound $c_0$ (using the peak density $n_0$). White arrows indicate the direction of the flow.  Simulation parameters are $N_\mathrm{s} = 4$, $\tau_\mathrm{s} = 5\,\mathrm{s}$, $R_0 = R_r$, $w_0 = 18 \, \mu\mathrm{m}$, and $U = \mu_0$.}
\label{fig:sims}
\end{figure}  
%

\begin{figure}[t!]
\includegraphics[width=\columnwidth]{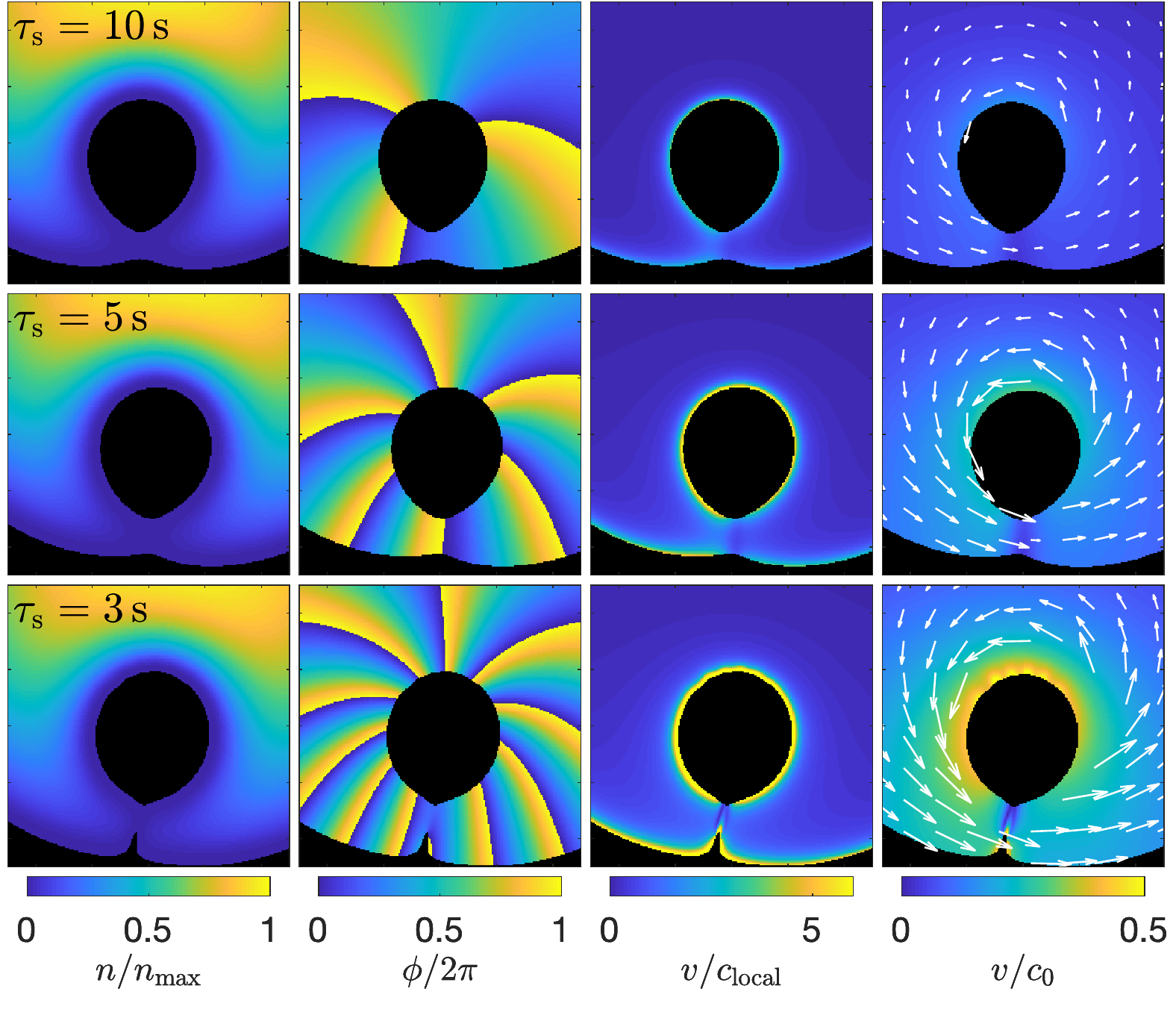}
\caption{Snapshots from 2D GPE simulations for varying $\tau_\mathrm{s}$ (as labeled). Snapshots show the timepoint $t_\mathrm{m} \sim 0.66 \tau_\mathrm{s}$ where the flow behind the beam merges with that ahead of the beam resulting in continuous superflow around the beam. Columns from left to right show zoomed in snapshots of (i) density, (ii) phase, (iii) the velocity profile scaled to the local speed of sound, and (iv) the velocity profile scaled to the bulk speed of sound $c_0$ (using the peak density $n_0$). White arrows indicate the direction of the flow.  Simulation parameters are $N_\mathrm{s} = 4$, $R_0 = R_r$, $w_0 = 18 \, \mu\mathrm{m}$, and $U = \mu_0$. Note the dark soliton appearing in the reconnecting flow for $\tau_\mathrm{s} = 3\,\mathrm{s}$.}
\label{fig:sims1}
\end{figure} 

To better understand the formation process of the multiply-quantized circulation state pinned to the spiraling beam, we performed numerical simulations using split-step Fourier evolution of the 2D Gross-Pitaevskii equation (GPE) \cite{Samson2016, Pethick2008}.  In particular, the numerical simulations enable visualization of the condensate's phase and velocity profiles throughout the spiral trajectory. Our method for numerical simulation of the 2D GPE follows that of Ref.~\cite{Samson2016}.  We reproduce the critical points here.  
We write the normalized 3D BEC wavefunction for our highly oblate BECs as the product of the axial $\Phi(z)$ and horizontal $\psi(x,y,t)$ wavefunctions
\begin{equation*}
\Psi(x,y,z,t) = \Phi(z) \psi(x,y,t),
\end{equation*}
where we assume any axial ($z$) dynamics are frozen due to the tighter confinement and approximate $\Phi(z)$ as a 1D Gaussian
\begin{equation*}
\Phi(z) = (\pi\ell_z^2)^{-1/4} e^{-z^2/2\ell_z^2}.
\end{equation*}
Indeed in our highly oblate geometry, vortex dynamics occur in the $\hat{x}$-$\hat{y}$ plane, and vortex-core excitations such as Kelvin modes are reduced \cite{Rooney2011}.
In the above expressions, $\ell_z = 0.51 \, \mu\mathrm{m}$ is an effective lengthscale for the axial width of the 3D BEC rather than the axial harmonic oscillator length $\ell_{\mathrm{HO},z} = \sqrt{\hbar/m\omega_z} = 1.1 \,\mu\mathrm{m}$.  We use the effective lengthscale together with an effective 2D atom number $N_\mathrm{2D} = 3.7 \times 10^5$ and an effective radial trap frequency $\tilde{\omega}_r = 0.84 \omega_r$ so that observables such as peak density $n_0$, radial Thomas-Fermi radius $R_r$, bulk chemical potential $\mu_0$, and the bulk speed of sound $c_0 = \sqrt{\mu_0/m} = 1800 \, \mu \mathrm{m/s}$ are consistent between the 2D simulations and the experimental parameters; see Ref~\cite{Samson2016}.  This allows us to better compare results from simulations and experiment.   

The 2D dynamics are modeled by the 2D GPE
\begin{equation*}
    (i-\gamma)\hbar\frac{\partial}{\partial t} \psi = \Big[-\frac{\hbar^2}{2 m}\nabla_{x,y}^2  + V_\mathrm{ht} + V_\mathrm{sb} + g_\mathrm{2D}\lvert \psi \rvert^2 \Big] \psi,
\end{equation*}
where $V_\mathrm{ht} = \frac{1}{2}m \tilde{\omega}_r^2(x^2 + y^2)$ is the harmonic trapping potential (in the absence of the stirring beam) and $g_\mathrm{2D} = \frac{4\pi\hbar^2 a_\mathrm{sc}}{m\sqrt{2\pi}\ell_z}$ is the effective 2D nonlinear interaction term, where $a_\mathrm{sc}$ is the atomic s-wave scattering length. 
For the GPE simulations we work in the reference frame of the BEC, so that the BEC is stationary and the stirring beam moves along the spiral trajectory. The time-dependent potential due to the blue-detuned stirring beam is
\begin{equation*}
    V_\mathrm{sb}(x,y,t) = U \exp \biggl\{-\frac{2}{w_0^2}\big[(x - x_\mathrm{s}(t))^2 + (y - y_\mathrm{s}(t))^2\big] \biggr\}
\end{equation*}
where $x_\mathrm{s}(t) \equiv r(t)\cos{\theta(t)}$ and $y_\mathrm{s}(t) \equiv r(t)\sin{\theta(t)}$ for $r(t)$ and $\theta(t)$ defined previously are the time-dependent positions of the moving stirring beam. $U$ is the maximum repulsive energy of the Gaussian stirring potential, and $w_0$ is the $1/e^2$ beam radius. We use imaginary-time propagation of the 2D GPE to generate the initial condition for the BEC in the full potential (harmonic trap + stirring beam).  We then model the dynamics using split-step evolution of the 2D GPE with $\gamma = 0.003$. Here we use $\gamma$ to phenomenologically account for finite-temperature effects such as damping due to the presence of the thermal component.  We use a spatial domain of 120 x 120 $\mu$m, and gridsize of either 512 x 512 (Figs.~\ref{fig:sims} and \ref{fig:sims1}) or 256 x 256 (Figs.~\ref{fig:number}-\ref{fig:fourbeams}). A gridsize of 512 x 512 corresponds to a spatial resolution of $0.23\,\mu\mathrm{m}$ on the order of the bulk healing length $\xi = \sqrt{\hbar^2/2m\mu_0} = 0.27 \, \mu\mathrm{m}$ and is therefore relevant when considering fluid merging dynamics especially at high merging speeds.

Figure~\ref{fig:sims} shows representative snapshots of the spiral method for $N_\mathrm{s} = 4$, $\tau_\mathrm{s} = 5\,\mathrm{s}$, $R_0 = R_r$, $w_0 = 18 \, \mu\mathrm{m}$, and $U = \mu_0$, corresponding to the spiral trajectory plotted in Fig.~\ref{fig:concept}. Columns from left to right show snapshots of 2D density $n(x,y,t) = N_\mathrm{2D} \vert \psi(x,y,t) \vert^2$, phase $\phi(x,y,t) = \mathrm{Arg}[\psi(x,y,t)]$, the velocity profile $v(x,y,t) = \frac{\hbar}{m}\nabla\phi(x,y,t)$, and a zoomed-in region of the velocity profile at the location of the stirring beam. The velocity profiles in the third column from the left are scaled to the effective 3D local speed of sound $c_\mathrm{local}(x,y,z = 0,t) = \sqrt{g \, n_\mathrm{eff,3D}(x,y,z=0,t) / m}$, where $n_\mathrm{eff,3D}(x,y,z=0,t) = \sqrt{\pi} \ell_z \, n(x,y,t)$ is the effective 3D atom density evaluated for $z = 0$. The velocity profiles in the fourth column are scaled to the bulk speed of sound $c_0 = \sqrt{\mu_0/m}$. White arrows indicate the direction of the superflow. 

The upper row shows the initial condition at $t/\tau_\mathrm{s} = 0$, with the spiral beam located at the edge of the BEC $r(t = 0) = R_r$.  As the velocity profiles show, the inward spiraling beam induces local superflow around the inner edge of the beam, i.e., the beam pushes the superfluid out of the way.  Eventually the flow behind the beam merges with that ahead of the beam such that the beam is fully surrounded by a continuous circular superflow.  
The flow around the beam at this merge timepoint $t_\mathrm{m}$ fixes the circulation pinned to the beam. 
Given the quantized nature of superfluid circulation, this in turn fixes the number of quantized vortices effectively pinned to the beam. 

Variations in the spiral parameters such as $N_\mathrm{s}$ and $\tau_\mathrm{s}$ change the speed of the stirring potential at the time of merging $v_\mathrm{s}(t = t_\mathrm{m})$, and in turn the speed of the fluid around the potential. As shown in Fig.~\ref{fig:sims1}, the flow speed increases with decreasing $\tau_\mathrm{s}$. Figure~\ref{fig:sims1} zooms in on the region around the spiral beam at the merge timepoint $t_\mathrm{m} \sim 0.66\tau_\mathrm{s}$ for varying $\tau_\mathrm{s}$ ($N_\mathrm{s} = 4$).  The superflow velocity profile becomes less smooth as the speed of the merging fluid increases.  In particular, for $\tau_\mathrm{s} = 3\,\mathrm{s}$ we start to see evidence of a dark soliton excitation forming \cite{Burger1999, Kevrekidis2019}, which then evolves into a single unpinned vortex. We note that for $\tau_\mathrm{s} = 3\,\mathrm{s}$, vortices depin from the stirring potential towards the end of the spiral trajectory.

\section{Choosing vortex winding number}\label{sec:winding}
\begin{figure}[t!]
\includegraphics[width=\columnwidth]{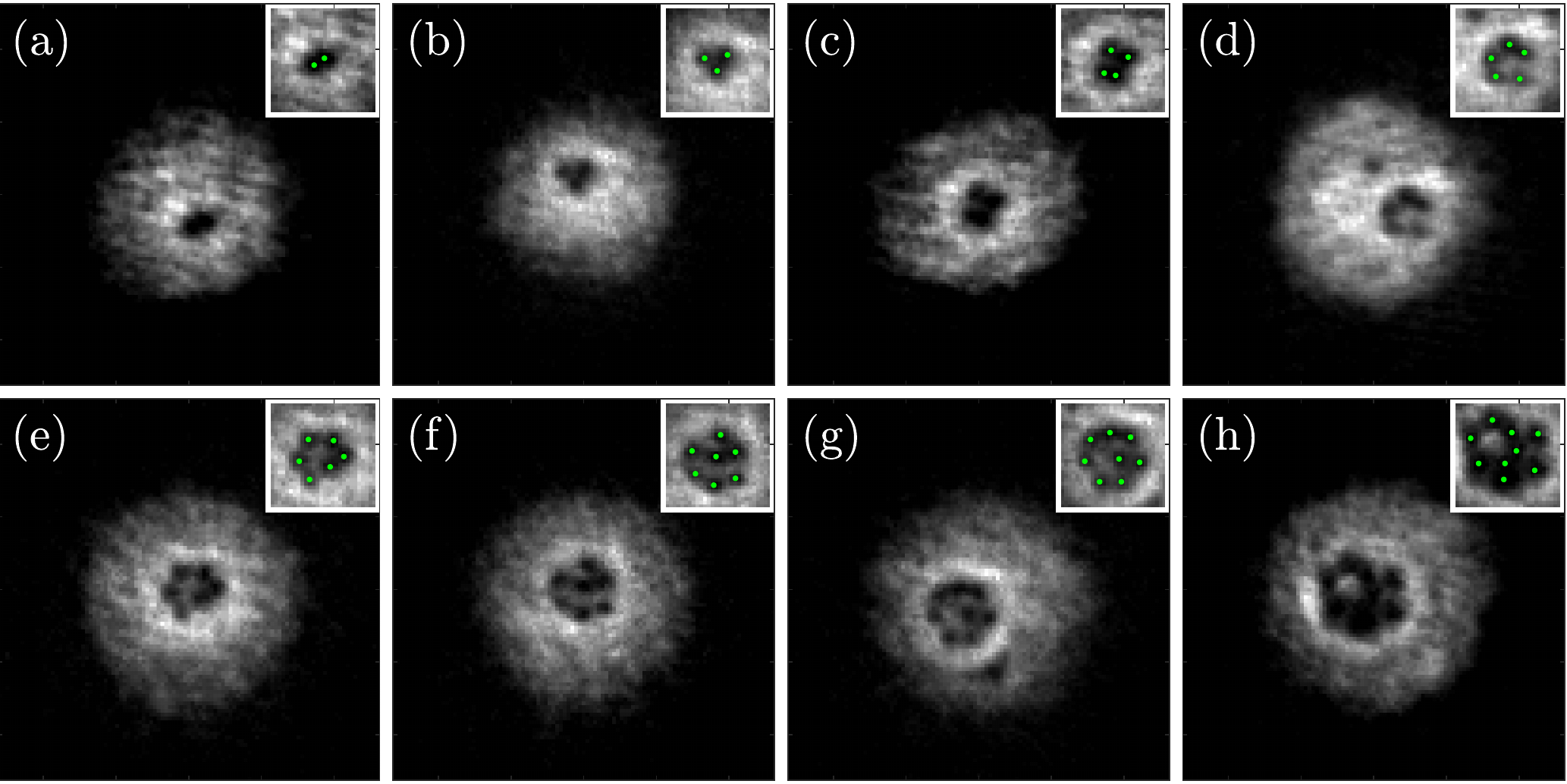}
\caption{Representative 200-$\mu$m-square axial absorption images of the BEC after ballistic expansion.  Expansion of the BEC starts $\sim 160 \, \mathrm{ms}$ after the pinning potential ramps off.  The pinned vortices start to dissociate and individual vortex cores can be resolved. (a)-(h) show vortex configurations for two to nine cores respectively.  Vortices are marked by green dots in the inset to guide the eye. Regular array configurations of the vortices can sometimes be observed for, i.e., (b) three, (c) four, and (e) six windings.} 
\label{fig:images}
\end{figure}

The vortex distributions shown in Fig.~\ref{fig:images} are representative of the range of multiply-quantized vortices created as we vary the spiral parameters $N_\mathrm{s}$ and $\tau_\mathrm{s}$. Since these individual vortices originate from the large pinned vortex configuration, we assume that they have the same sign of circulation, which is determined by the direction of the spiral.  We note that regular array configurations of the vortices can sometimes be observed, e.g., for (b) three, (c) four, and (e) six windings, as reported in previous numerical simulations \cite{Capuzzi2009, Cawte2021}.

\begin{figure}[t!]
\includegraphics[width=\columnwidth]{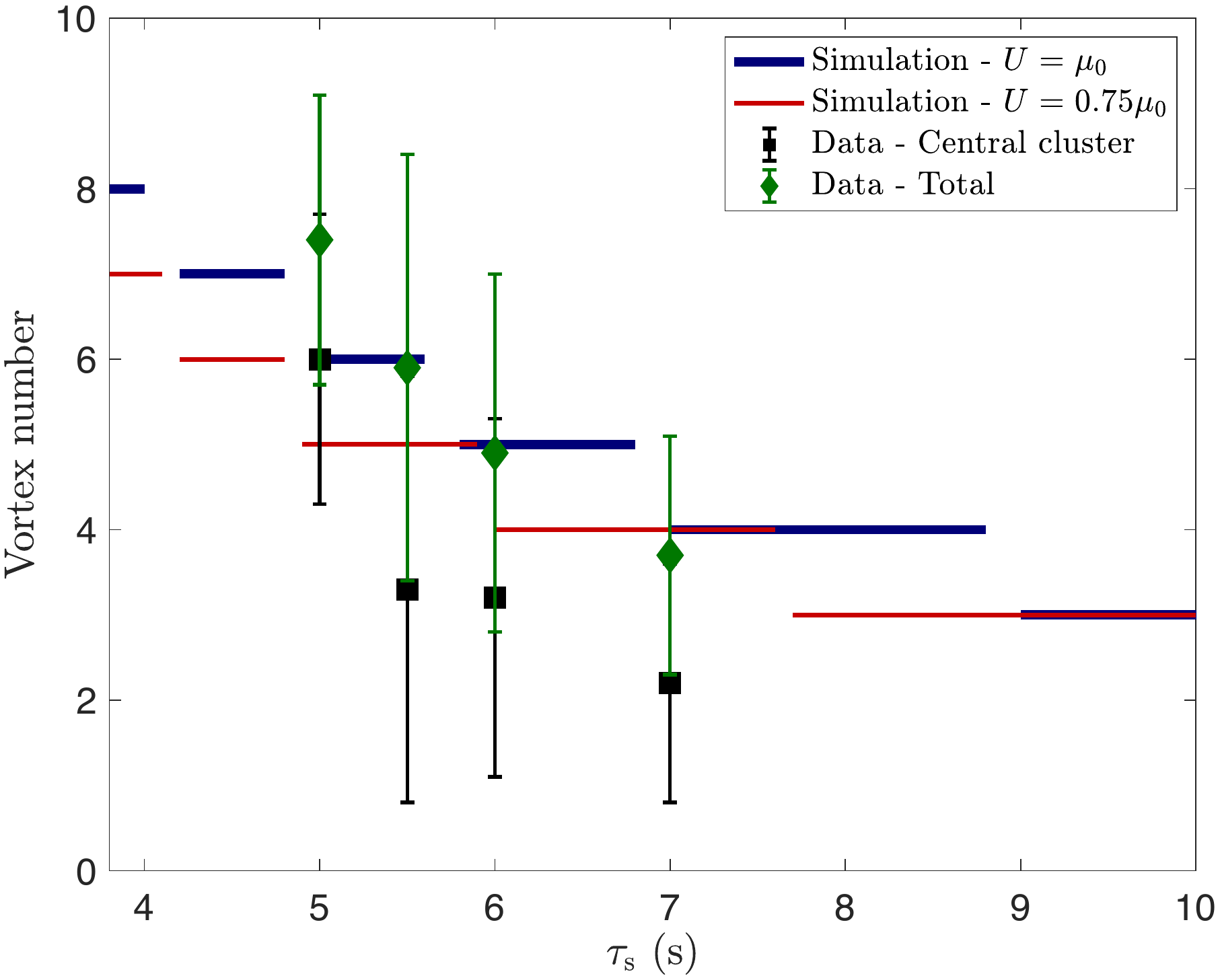}
\caption{Average number of vortices observed after ramping off the pinning potential as a function of trajectory duration $\tau_\mathrm{s}$, for fixed $N_\mathrm{s} = 4$ and beamwaist $w_0 = 18 \, \mu\mathrm{m}$.  Black squares and green diamonds correspond to experimental data, while solid blue and red lines are numerical results from the 2D GPE simulations.  Black squares show the average number of cores in the central cluster, whereas green diamonds show the average total number of vortex cores.  Error bars indicate the standard deviation. Solid blue lines correspond to a spiraling potential $U = \mu_0$, and solid red lines correspond to $U = 0.75\mu_0$.  Both experiment and numerics show that the number of vortices generated and pinned using our method decreases as the optical potential moves more slowly through the condensate.}
\label{fig:number}
\end{figure}

Ultimately we want to be able to control the exact winding number and placement of the vortex cluster.  To this end, 
we further explore the relationship between stirring speed and the number of pinned vortices, by varying the time duration of the spiral trajectory. 
In Fig.~\ref{fig:number} we plot the number of vortices observed after ramping off the pinning potential as a function of the spiral trajectory duration $\tau_\mathrm{s}$ for a fixed $N_\mathrm{s} = 4$.  Black squares and green diamonds show experimental data; black squares show the average number of cores in the central cluster, whereas green diamonds show the average total number of vortex cores regardless of their position in the condensate.  Error bars indicate the standard deviation.  The solid blue lines in Fig.~\ref{fig:number} correspond to simulations using $U = \mu_0$, and the solid red lines correspond to $U = 0.75\mu_0$.  Results from our 2D GPE simulations are qualitatively consistent with experiment, with longer spiral times resulting in fewer pinned vortices. 
We note that we tend to observe 1-2 vortices outside of the central cluster even with the use of a hold prior to ramping off the blue-detuned pinning beam.  These are likely vortices that have left the pinning potential prematurely. Our specific spiral trajectory stops abruptly at the end of the spiral which may dislodge some vortices. Refinement of this trajectory is the subject of future work. 

Both experiment and numerics show that the winding number of the pinned superflow decreases with the trajectory duration. At the fastest spiral trajectory ($\tau_\mathrm{s} = 5$ s), the mean winding number generated by our method is $\sim \! 6 \text{ to } 7.4$, while at the slowest ($\tau_\mathrm{s} = 7$ s), the mean is $\sim \! 2.2 \text{ to } 3.7$ vortex cores.  The highest winding number that we were able to observe at a single occurrence was 11.  
The number of vortices that can be stably pinned is limited by the radius of the pinning potential and the angular-rotation frequency of the system \cite{Simula2002}.  In some of the condensates, vortices may have prematurely left the pinning potential either as the beam comes to a halt or during the beam ramp off. These vortices can be seen as vortex cores outside the regular cluster of dissociating vortices [see e.g., Fig.~\ref{fig:exp}(d) and Fig.~\ref{fig:images}(d)].  

The GPE simulations indicate that the spiral method should be fairly robust.  A $\pm10\%$ change in atom number $N$, initial spiral radius $R_0$, beamwaist $w_0$, height of the blue-detuned potential $U$, or the aspect ratio of the spiral resulted in variation of the number of pinned cores by $\pm1$.  We found similar variation when shifting the initial beam positions $x_0$ and $y_0$ by $\pm 5~\mu\mathrm{m}$ out of a Thomas-Fermi radius $R_r =\sim 50~\mu\mathrm{m}$.  In our experiments, the largest impediment to reproducibility was drift in the axial alignment of the blue-detuned potential with respect to the center of the BEC. We experienced both a long-time-scale drift that could be accounted for by periodically centering the beam to the final BEC position, and also shot-to-shot fluctuations due to the BEC receiving an arbitrary kick earlier in the evaporation sequence (likely as a magnetic field turns off). In future, the shot-to-shot fluctuations may be mitigated by ramping on the blue-detuned beam prior to the final stages of evaporation. With this in mind, optical traps using digital micro-mirror devices \cite{Gauthier2016} where the stirring beam and the optical trap are generated by the same laser beam may be best suited for future implementation of our technique.

\section{Extensions to multiple beams}\label{sec:twobeams}
%
\begin{figure}[t!]
\includegraphics[width=\columnwidth]{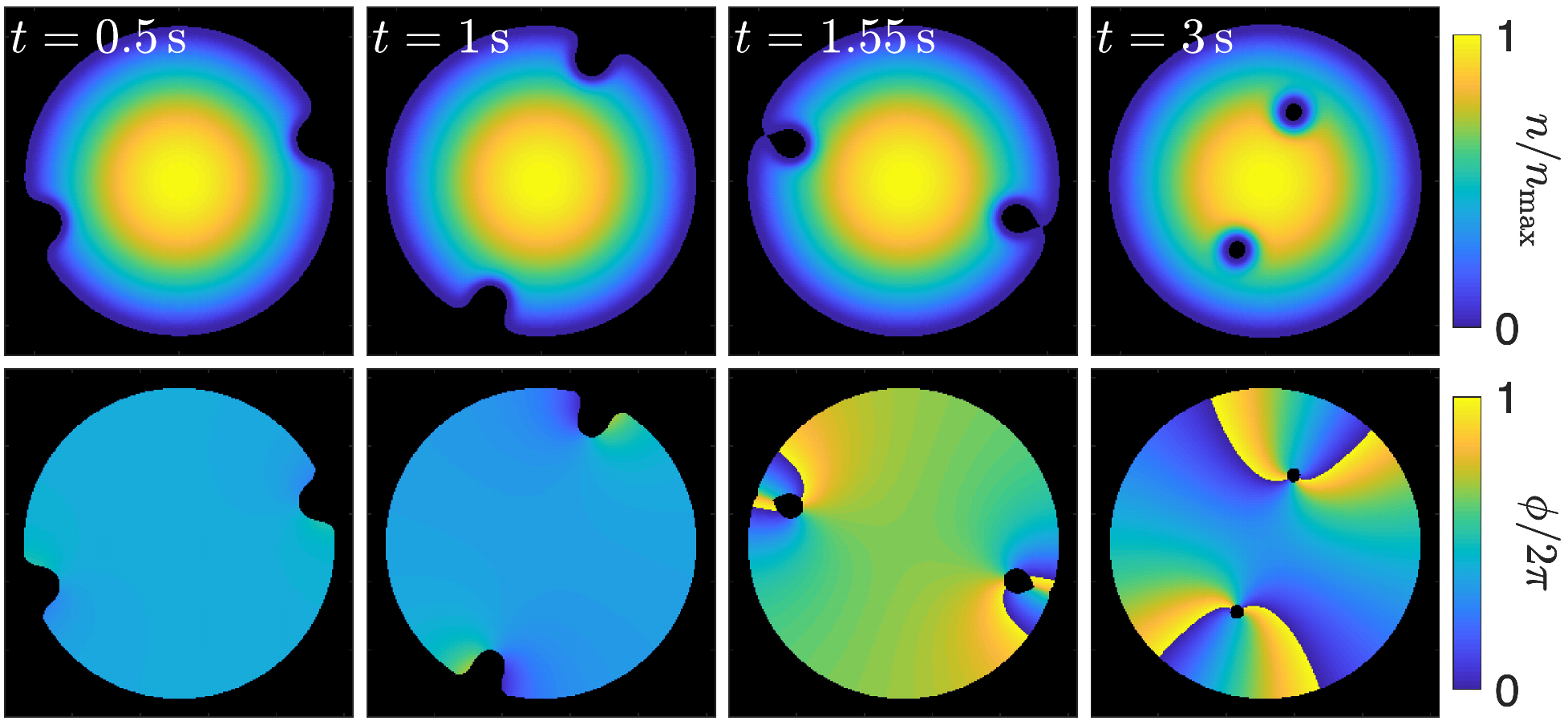}
\caption{GPE simulation for two beams both spiraling counter-clockwise.  Simulation parameters are $N_\mathrm{s} = 3$, $\tau_\mathrm{s} = 4\,\mathrm{s}$, $w_0 = 8 \, \mu\mathrm{m}$, and $U = \mu_0$.}
\label{fig:twobeams2}
\end{figure}

We use the 2D GPE simulations to explore the potential for extending the spiral technique to multiple beams.  Here we have decreased the beamwaists to $w_0 = 8 \, \mu\mathrm{m}$ largely to fit more beams within the finite size of the condensate $R_r \sim 50 \, \mu\mathrm{m}$. Reducing the beamwaist does result in a reduction of beam speed at the time when the beam fully enters the condensate $v_\mathrm{s}(t = t_\mathrm{m})$, thus resulting in fewer pinned cores for a given $\tau_\mathrm{s}$.  However, reducing the beamwaist does not fundamentally alter the spiral technique.  For Fig.~\ref{fig:twobeams2} we simply add a second spiral beam on the opposite side of the condensate.  Both beams then follow the same trajectory just $180^{\circ}$ out of phase.  Both beams spiral in a CCW direction with the same spiral parameters, and we find each beam generates a doubly-quantized pinned circulation with the same sign of circulation.   

\begin{figure}[t!]
\includegraphics[width=\columnwidth]{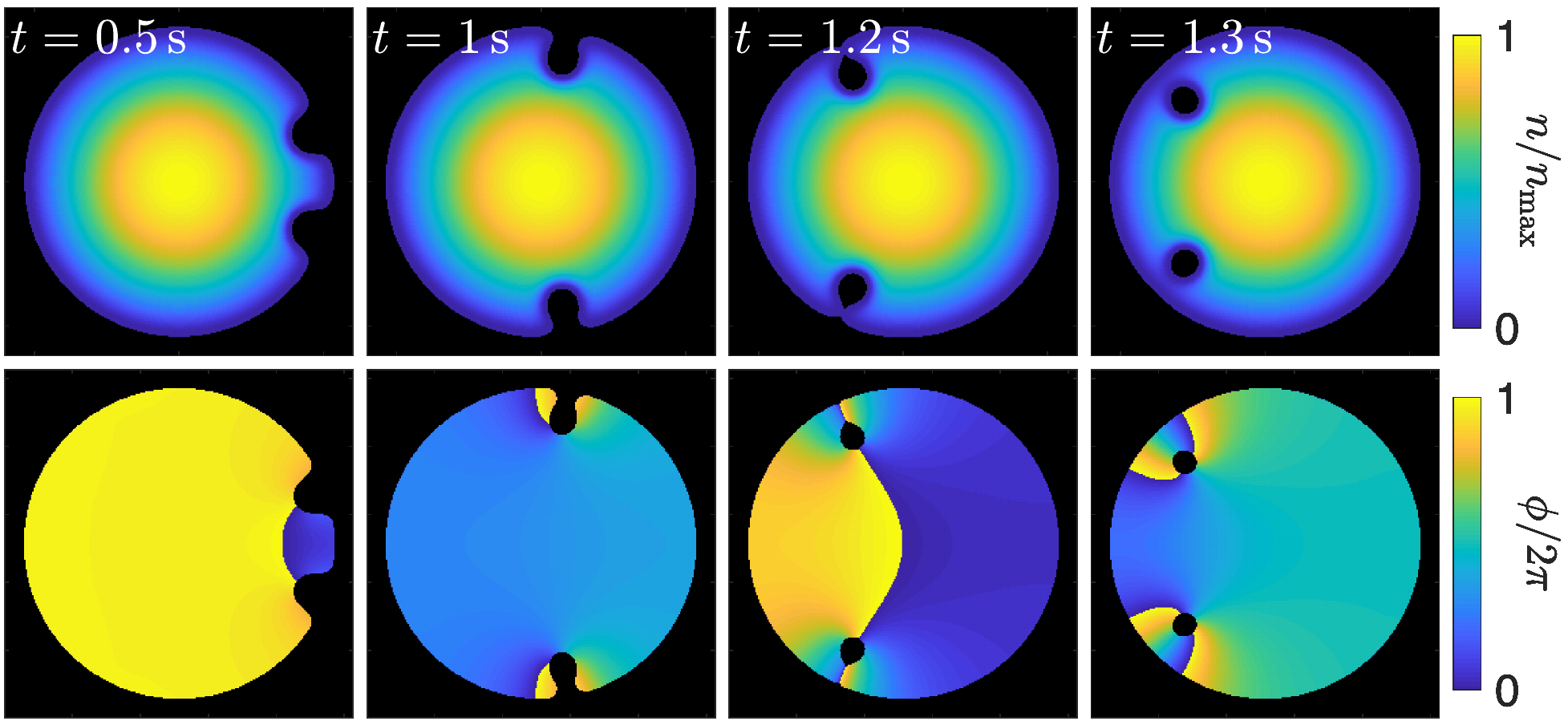}
\caption{GPE simulation for two beams, one spiraling clockwise and one spiraling counter-clockwise.  Simulation parameters are $N_\mathrm{s} = 2$, $\tau_\mathrm{s} = 3\,\mathrm{s}$, $w_0 = 8 \, \mu\mathrm{m}$, and $U = \mu_0$.}
\label{fig:twobeams1}
\end{figure}

The unique feature of our spiral technique is that each beam creates and pins vortices all of the same winding number, as opposed to direct nucleation of vortex dipoles \cite{Inouye2001, Neely2010, Samson2016}.  However, as shown in Fig.~\ref{fig:twobeams1} we can engineer a scenario where we employ two beams, one spiraling CCW and one spiraling CW.  In this scenario, we end up with two regions of opposite-signed multi-quanta circulation.  We note that this scenario does require modifying the spiral trajectory after the vortices have been generated to avoid a collision of the pinning potentials and subsequent depinning or annihilation of the vortex cores.
%

\begin{figure}[t!]
\includegraphics[width=\columnwidth]{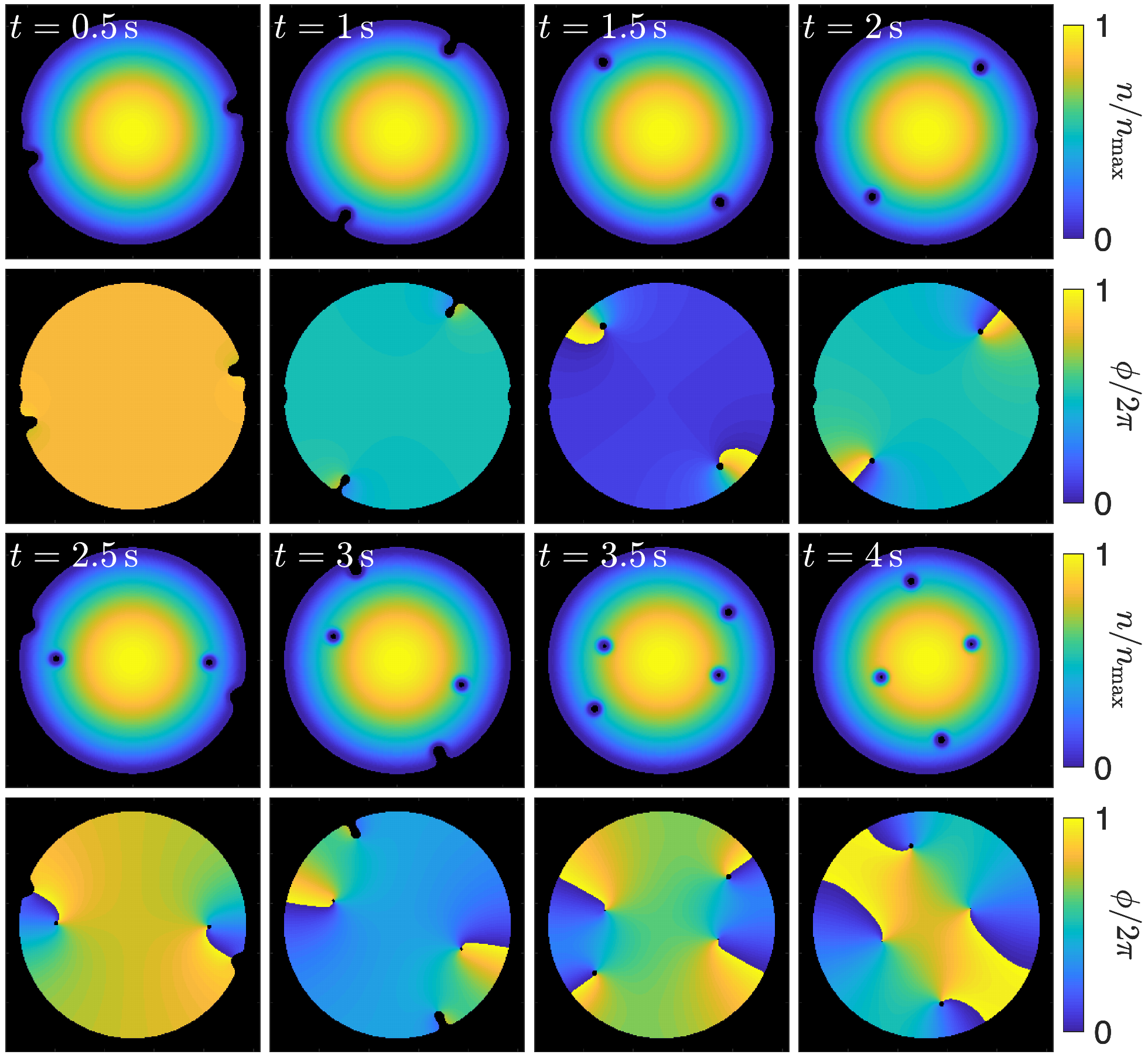}
\caption{GPE simulation for four beams.  First one pair of beams spirals counter-clockwise.  After 2\,s of evolution, the second pair of beams begins to spiral clockwise.  Simulation parameters are $N_\mathrm{s} = 4$, $\tau_\mathrm{s} = 5\,\mathrm{s}$, $w_0 = 4 \, \mu\mathrm{m}$, and $U = \mu_0$.}
\label{fig:fourbeams}
\end{figure}

In Fig.~\ref{fig:fourbeams} we extend the technique further to four beams each with a beamwaist of $w_0 = 4 \, \mu\mathrm{m}$. In this final scenario, first one pair of beams spirals CCW, similar to that shown in Fig.~\ref{fig:twobeams2}. After 2\,s of evolution, a second pair of beams begins to spiral CW.  The net result is four pinning potentials within the BEC, each guiding a singly-quantized vortex.  Here the vortices pinned by the second pair of beams have the opposite sign from the vortices pinned by the first pair of beams.  

The technique can be readily extended to trapping potentials with hard walls and flat bottoms.
However, for smaller beams it is crucial to be able to control the relative position of the trap and the beam.  In particular, it is detrimental if the beams dip in and out of the condensate rather than spiraling smoothly inwards.  Therefore extensions of this technique are likely best suited to a DMD-style optical potential \cite{Reeves2020, Gauthier2016}. In this scenario both trap and stirring beam are generated from the same laser beam, and one can take advantage of common-mode noise rejection.

\section{Conclusion}\label{sec:conclusion}
We have demonstrated a method for deterministically generating multiply-quantized circular superflow in a highly oblate $^{87}$Rb Bose-Einstein condensate.  Using an optical potential moving in a spiral trajectory toward the center of the condensate, we generate multiply-quantized circulation with winding numbers as high as 11.  By then ramping off the optical potential, a cluster of quantized vortices can be released into the condensate.  By changing the rate at which the beam moves along this trajectory, we can selectively control the vorticity introduced into the BEC. Our spiral method presents a reliable source of pinned circulation and quantized vortices for superfluid and BEC studies that require superfluid circulation with high winding numbers, and extends readily to multiple stirring beams and hard-wall trapping potentials.  This will prove useful for extending experimental studies regarding stability and dissociation of multiply-charged vortices having higher circulation quanta, as well as experimental studies of 2D turbulence, point vortex dynamics, and analog cosmology. 

\begin{acknowledgments}
We thank Ewan Wright for helpful discussions, and acknowledge funding from the US National Science Foundation, grant number PHY-1205713. K.E.W. acknowledges support from the US Department of Energy Office of Science Graduate Fellowship Program, administered by ORISE-ORAU under Contract No. DE-AC05-06OR23100, and funding from the Royal Society University Research Fellowship. 
\end{acknowledgments}

%

\end{document}